\documentclass[RNAAS]{aastex62}

\graphicspath{{./}{figures/}}

\begin{document}

\title{TESS Extended Mission 10-Minute Cadence Retains Nyquist Aliases}

\correspondingauthor{Keaton J.\ Bell}
\email{keatonb@uw.edu}

\author{Keaton J.\ Bell}
\affiliation{DIRAC Institute, Department of Astronomy, University of Washington, Seattle, WA-98195, USA}
\affiliation{NSF Astronomy and Astrophysics Postdoctoral Fellow and DIRAC Fellow}


\section{}

During its two-year prime mission, the Transiting Exoplanet Survey Satellite \citep[TESS;][]{Ricker2014} is obtaining full-frame images with a regular 30-minute cadence in a sequence of 26 sectors that cover a combined 85\% of the sky. While its primary science case is to discover new exoplanets transiting nearby stars, TESS data are superb for studying many types of stellar variability, with the number of publications using TESS data for other areas of astrophysics keeping pace with exoplanet papers.\footnote{\url{https://heasarc.gsfc.nasa.gov/docs/tess/publications.html\#breakdown-by-subject}}
Following the conclusion of its prime mission in July 2020, TESS will revisit the sky in an extended mission that records full-frame images at a faster ten-minute cadence.  
In this note, I demonstrate that choosing a large submultiple of the original exposure times for the new cadence limits the synergy between prime and extended TESS mission data since both sampling rates produce many of the same Nyquist aliases. Adjusting the extended mission exposure time by as little as one second would largely resolve Nyquist ambiguities in the combined TESS data set.

Recording any periodic signal with a regular time sampling of $\Delta t$ will produce an infinite number of identical alias peaks in a periodogram that are reflected off of the Nyquist frequency, $f_\mathrm{Nyq} = (2\Delta t)^{-1}$.
This means that there are an infinite number of candidate frequency solutions that do an equally good job of explaining the regularly sampled signal. Without a physical argument for choosing one of these candidates over the others, analyses of TESS data are ambiguous at best and inaccurate at worst.
In these cases, it is typically necessary to obtain follow-up observations with a different cadence to resolve which peaks correspond to intrinsic frequencies \citep[e.g.,][]{Bell2017}.
This strains observing resources, limits results to researchers with abundant telescope access, and is impractical for the millions of stars that TESS is observing.
TESS itself could eliminate the need for such follow-up if it were to carry out its extended mission observations with a sample spacing that is not precisely a small submultiple of the original cadence.

\begin{figure}[t]
\begin{center}
\includegraphics[scale=0.85,angle=0]{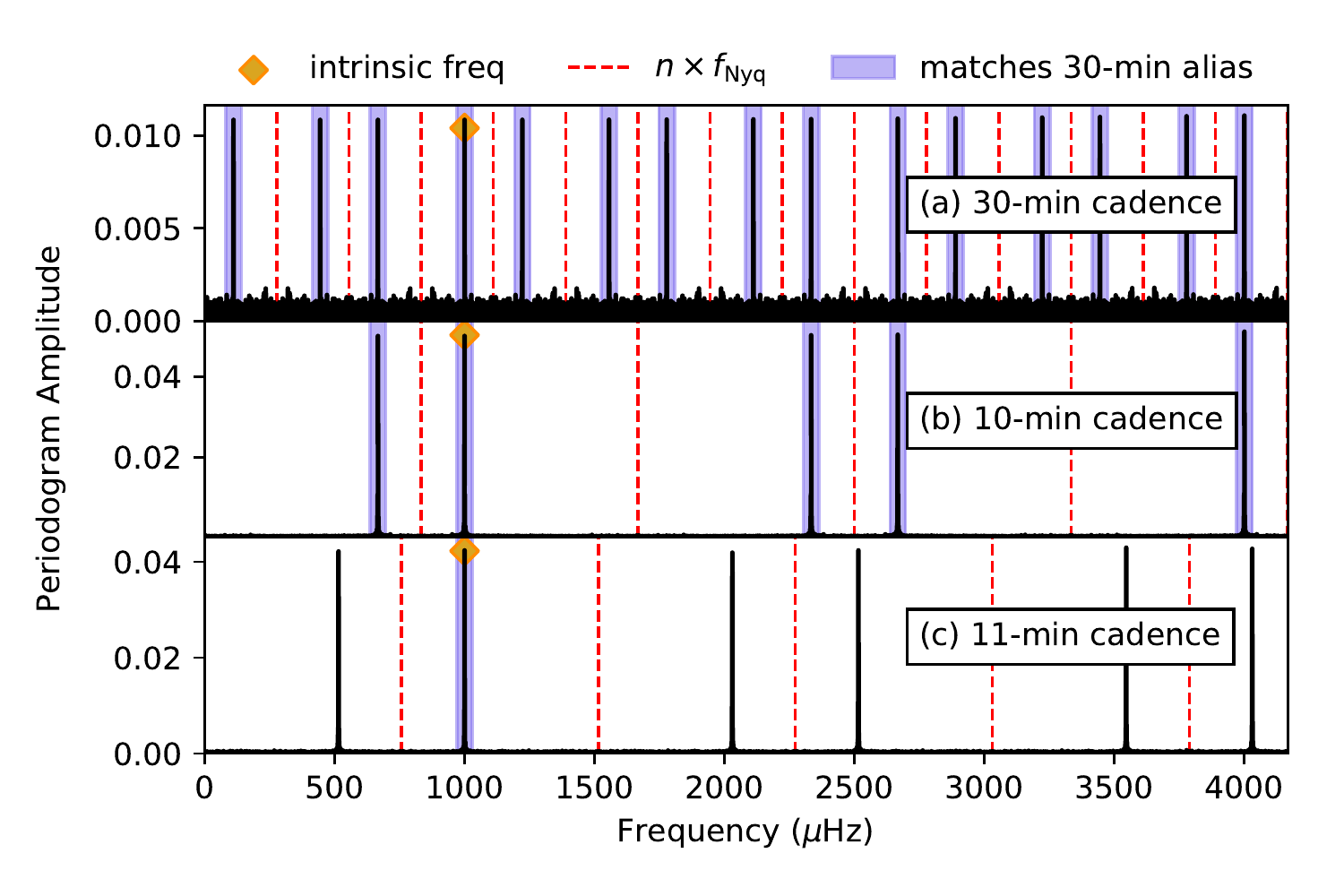}
\caption{Periodograms of a simulated light curve with a single 1000\,$\mu$Hz signal for three sampling rates: (a) the TESS prime mission's 30-minute cadence; (b) TESS's 10-minute extended mission cadence; and (c) a hypothetical 11-minute cadence. The dashed lines indicate multiples of the Nyquist frequencies that aliases are reflected across, the orange diamonds mark the peak corresponding to the intrinsic 1000\,$\mu$Hz underlying signal, and all peaks that are precisely coincident with aliases from the 30-minute TESS light curve are highlighted in blue.
\label{fig:1}}
\end{center}
\end{figure}

Figure~\ref{fig:1} depicts the Lomb-Scargle periodograms of simulated light curves sampled at three rates: (a) once every 30 minutes; (b) once every 10 minutes; and (c) once every 11 minutes. The underlying source exhibits a single signal with an intrinsic frequency of 1000\,$\mu$Hz, resulting in an infinite set of aliases in each periodogram.  Because the spacing between samples in (b) is a submultiple of the sampling of (a), all aliases in periodogram (b) are exactly coincident with aliases in (a), and the solution from the combined data set remains ambiguous. However, since the sampling of (c) is not so simply related to (a), the peak at the intrinsic frequency is easily identified as the only one that appears at the same location in both periodograms (a) and (c).

The same conclusion holds for periodic signals that are not strictly sinusoidal, since all harmonics in the periodogram are also aliased to the same frequencies for 30- and 10-minute cadences. 
The ratio of amplitude reduction from smoothing by 30- compared to 10-minute integration times is also the same at the locations of all aliases.  
In rare cases when signals that are known to be exceedingly coherent, the small deviations from regular sampling introduced by barycentric corrections for spacecraft motion may be exploited to identify an intrinsic signal from among its aliases, yet the leverage from this effect is much smaller for TESS than it was for \emph{Kepler} \citep{Murphy2015}.

The sampling of $\Delta t=11$\,min in Figure~\ref{fig:1} was chosen for dramatic visual effect; in fact, considering that 27 days of TESS observations per sector yields a frequency resolution of $\approx 0.43\,\mu$Hz, an adjustment from the planned 10-minute exposure time of just less than one second would be sufficient to 
shift the Nyquist frequency and the aliases it creates into different frequency bins between prime and extended mission periodograms.\footnote{Signal frequencies are typically measured to better precision than the frequency resolution \citep{Montgomery1999}.} The TESS hardware restricts exposures to multiples of 20 seconds (R.~Vanderspek, private communication), and an adjustment by this larger amount would further help resolve Nyquist ambiguities for stochastic signals with power that spans multiple frequency bins \citep[e.g.,][]{Chaplin2014}, with minimal effect on the temporal resolution of the light curves.
Varying the phase coverage of signals between prime and extended missions in this way also stands to benefit exoplanet studies by improving the combined sampling of transit profiles. 
If it is too late to make a cadence adjustment for TESS Cycle 3 that begins the extended mission in July 2020, I urge that later cycles that revisit the same areas of sky be carried out with a sampling designed to mitigate the confusion from Nyquist aliases in the final TESS legacy data set.

\acknowledgments

This material is based upon work supported by the National Science Foundation under Award No.\ AST-1903828. Thanks to E.~Bellm and J.~R.~A.~Davenport for feedback.

\end{document}